\documentclass[11pt, oneside]{article}   	
\usepackage{geometry}                		
\usepackage{graphicx}				
									
\usepackage{amssymb}
\usepackage{amsmath}
\usepackage{bm}					
\usepackage[symbol]{footmisc}

\title{Feynman's Relativistic Cut-Off for Quantum Electrodynamics}
\author{Hari Chapagain}

\begin{document}
\maketitle
%\noindent
\abstract {Feynman's modification to electrodynamics and its application to the calculation of self-energy of a free spin-$\frac{1}{2}$ particle, appearing in his 1948 \emph{Physical Review} paper,$^{\scriptsize{\cite{FeynmanQuantum}}}$ is shown to be applicable for the self-energy calculation of a free spin-0 particle as well. Feynman's modification to electrodynamics is shown to be equivalent to a Hamiltonian approach developed by Podolsky.$^{\scriptsize{\cite{PodolskySchwed}}}$}
\section{Introduction}
In the early days of QED, around 1930s, calculations of various electrodynamic processes were plagued by infinities. During the Shelter Island conference, in 1947, Kramers made the suggestion that certain infinities, such as the self-energy, could be absorbed into the mass of the electron. On a train ride from New York to Schenectady after the conference, Bethe did a nonrelativistic calculation of the 2S-2P level shift for hydrogen employing the mass renormalization idea suggested by Kramers. He got a logarithmic divergence for the shift because in the non-relativistic theory the electron self-energy is linearly divergent. He suspected that a relativistic treatment of the electron would yield a finite result for the level shift because the self-energy would only be logarithmically divergent.$^{\scriptsize{\cite{Schweber}}}$\\
\\
Returning to Cornell, after the Shelter Island conference, Bethe gave a lecture on the subject and suggested that if there were any modifications (relativistic) to electrodynamics that would render the self-energy calculation finite (even if the resulting theory were unphysical) then carrying out the mass renormalization and level-shift calculation would be easy and unambiguous.$^{\scriptsize{\cite{FeynmanNobel}}}$ \\
\\
After Bethe's lecture Feynman approached Bethe and said he could do it. During his graduate years in Princeton Feynman had developed, in collaboration with his advisor John Wheeler, a version of electrodynamics called the ``Wheeler-Feynman absorber theory". He had studied in detail every which way to modify electrodynamics.$^{\scriptsize{\cite{FeynmanNobel}}}$ Feynman's modification was to change the Dirac-delta $\delta$ appearing in the Fokker action \eqref{FokkerAction} for classical electrodynamics by a narrow peaked function $f$.$^{\scriptsize{\cite{FeynmanClassical}}}$
\begin{equation}\label{FokkerAction}
	S= \sum_{a} {m_{a} \int \sqrt{da_{\mu} da^{\mu}}} + \sum_{\underset{a,b}{a\neq b}}{ e_{a}e_{b} \int  \int \delta(s^{2}_{ab}) da_{\mu}db^{\mu}}
\end{equation}
\\
Feynman's modification to quantum mechanical perturbation calculation of self-energy was to first write the term $d\bm{k}/k$ (appearing in the calculations) equivalently as $2\int \delta(\omega^2 - k^2) d\omega d\bm{k}$, where integration over $\omega$ is taken only for positive values, and then to replace the $\delta(\omega^2 - k^2)$ by
\begin{equation}\label{FeynmanMod}
	\int_{0}^{\infty} \left( \delta(\omega^2 - k^2) - \delta(\omega^2 - k^2 - \lambda^2) \right) G(\lambda) d\lambda,
\end{equation}
where $G(\lambda)$ is some smooth function such that $\int_{0}^{\infty} G(\lambda)d\lambda = 1$.
This is equivalent to a choice of $f$ shown in eq. \eqref{Ffunction}.$^{\scriptsize{\cite{FeynmanSpaceTime}}}$
Feynman applied this relativistic cut-off procedure to a free electron (spin-1/2 case) interacting with the electromagnetic field and showed that the self-energy of a free electron does indeed have the form of a relativistic mass correction. This was shown in his 1948 \emph{Physical Review} paper titled ``Relativistic Cut-Off for Quantum Electrodynamics".$^{\scriptsize{\cite{FeynmanQuantum}}}$\\
\\
Feynman's method of doing quantum mechanics was to start from an action integral, e.g. Fokker action \eqref{FokkerAction}, and then using his ``path integral" approach get results, agreeing with standard quantum theory, directly without the gymnastics of the usual Hamiltonian approach.
The choice of $f$ shown in eq. \eqref{Ffunction} is equivalent, in the Hamiltonian scheme, to adding an extra term $-\frac{1}{2}a^{2} \frac{\partial F^{\mu\nu}}{\partial x^{\nu}}\frac{\partial F_{\mu\gamma}}{\partial x_{\gamma}}$ to the usual Lagrangian density of the electromagnetic field.$^{\scriptsize{\cite{PodolskySchwed}}}$ This equivalence of the choice of $f$ shown in eq. \eqref{Ffunction} and the Hamiltonian scheme is shown in section \ref{FeynmanHamiltonian} of this paper. \\ 
\\
In 1948, Dyson had calculated the self-energy of a free electron while neglecting its spin (essentially a spin-0 case).$^{\scriptsize{\cite{Dyson}}}$ This was published before Feynman's 1948 \emph{Physical Review} paper, so the modification suggested by Feynman had not been applied for the spin-0 case. The self-energy calculation for spin-0 case is algebraically simpler, compared to the spin-$\frac{1}{2}$ case, because of the absence of spin degree of freedom. In section \ref{section3} of this paper, Feynman's modification \eqref{FeynmanMod} is applied to the spin-0 case using the Hamiltonian scheme.\\
\\
Carrying out the integrals \eqref{FeynmanSelfEnergy} involved in the self-energy calculation of spin-$\frac{1}{2}$ case is rather tedious and involves various substitutions and integration by parts to reduce to elementary integrals.$^{\scriptsize{\cite{FeynmanQuantum}}}$ In this paper, the terms obtained in the self-energy calculation of spin-0 case are reduced to similar terms from the spin-$\frac{1}{2}$ case. Thus the relativistic cut-off procedure carried out by Feynman for spin-$\frac{1}{2}$ case can readily be applied to the spin-0 case as well.\\
\\
Carrying out the self-energy calculation of a free electron using non-covariant perturbation method results in some terms that don't have the form of a relativistic mass correction. On account of the result being formally infinite, self-energy cannot be unambiguously mass renormalized using non-covariant ``old-fashioned perturbation" method. This struggle of the old-fashioned perturbation theory was remedied by the development of QED calculation techniques that made Lorentz invariance manifest throughout the calculations.\\
\\
In a theory free of infinities, calculation of self-energy of a free electron (even using the old-fashioned perturbation method) would result in terms that have the form of a relativistic mass correction, which could then be absorbed into the mass of the electron to carry out the mass renormalization procedure.
%%%%%%%%%%%%%%%%%%%%%%%%%%%%%%%%%%%%%%%%%%%%%%%%%%%%%%%%%
%%%%%%%%%%%%%%%%%%%%%%%%%%%%%%%%%%%%%%%%%%%%%%%%%%%%%%%%%
\section{Feynman's modification in the Hamiltonian method}\label{FeynmanHamiltonian}
During the 1940s several physicists were investigating different modifications to classical electrodynamics to solve the self-energy problem. Podolsky \& Schwed published$^{\scriptsize{\cite{PodolskySchwed}}}$ one such modification where they simply added an extra term $-\frac{1}{2}a^{2} \frac{\partial F^{\mu\nu}}{\partial x^{\nu}}\frac{\partial F_{\mu\gamma}}{\partial x_{\gamma}}$ to the usual Lagrangian density and were able to get a modified electrodynamics that was free from infinities due to a point source. In this modified electrodynamics the four-vector potential $A_{\mu}(x)$ satisfies the following relation\footnote{In Gaussian units.}
\begin{equation}\label{AvectorEqPodSch}
	a^{2}\Box^{2} A_{\mu}+ \Box A_{\mu}= 4\pi j_{\mu},
\end{equation}
where $\Box=\partial^{\mu}\partial_{\mu}$. The usual relation is obtained by setting $a=0$.\\
\\
To carry out the perturbation calculation one needs the mode expansion of the free fields $A_{\mu}$ satisfying
\begin{equation}\label{AvectorEqNoSource}
	a^{2}\Box^{2} A_{\mu}+ \Box A_{\mu}= 0.
\end{equation}
It can be shown that $A_{\mu}= A^{(1)}_{\mu} - A^{(2)}_{\mu}$, such that $\Box A^{(1)}_{\mu}= 0$ and $a^{2} \Box A^{(2)}_{\mu} + A^{(2)}_{\mu} = 0$, satisfies eq. \eqref{AvectorEqNoSource}. Thus the mode expansion of $A_{\mu}$ can be expressed as the difference between the mode expansions of the fields $A^{(1)}_{\mu}$ and $A^{(2)}_{\mu}$. $A^{(1)}_{\mu}$ is the usual massless photon field with $E_{p}= \bm{p}$ as its dispersion relation whereas $A^{(2)}_{\mu}$ describes a massive photon field with $E_{p}= \sqrt{\bm{p}^{2} + 1/a^{2}}$ as its dispersion relation. Feynman's modification \eqref{FeynmanMod} effects the same correction with $G(\lambda)= \delta(\lambda - \lambda_{0})$.\\
\\
Feynman's modification to electrodynamics was effected by replacing the Dirac-delta $\delta$ appearing in the Fokker action \eqref{FokkerAction} by an appropriate function $f$. A convenient choice for $f$ made by Feynman was\footnote[4]{This expression was given incorrectly in \cite{FeynmanQuantum} but later corrected in \cite{FeynmanSpaceTime}.}
\begin{equation}\label{Ffunction}
	f= \frac{1}{2\pi^{2}}\int \left(\delta(k_{0}^2 - k^2)- \delta(k_{0}^2 - k^2 - \lambda^{2})\right)\sin(k_{0}|t|)\cos(\bm{k}\cdot \bm{x})\Theta(k_{0}) G(\lambda)\ dk_{0} d\bm{k}\ d\lambda.
\end{equation}
The four-vector potential $A_{\mu}(x)$ due to particle ``$b$" (defined originally with a $\delta$ function) becomes
\begin{equation}\label{modifiedA}
	A_{\mu}(x) = \int f(s_{xb}^{2}) db_{\mu},
\end{equation}
where $s_{xb}^{2}=(x-b)_{\mu}(x-b)^{\mu}$ and $b_{\mu}$ are the coordinates of $b$. It can be shown by direct substitution that for $G(\lambda)=\delta(\lambda - \lambda_{0})$
\begin{equation}\label{AvectorEqFeyn}
	\Box^{2} A_{\mu}+ \lambda_{0}^{2}\ \Box A_{\mu}= \lambda_{0}^{2}\ 4\pi j_{\mu},
\end{equation}
where $j_{\mu}= \int \Pi_{\nu}\delta(x_{\nu}-b_{\nu}) db_{\mu}$. This is same as eq. \eqref{AvectorEqPodSch} with $a=1/\lambda_{0}$.\\
\\
Feynman's modification can thus be carried out in the Hamiltonian scheme as well.
%%%%%%%%%%%%%%%%%%%%%%%%%%%%%%%%%%%%%%%%%%%%%%%%%%%%%%%%%
%%%%%%%%%%%%%%%%%%%%%%%%%%%%%%%%%%%%%%%%%%%%%%%%%%%%%%%%%
\section{Self-energy for spin-0 case}\label{section3}
The self-energy $(\Delta E_{p}\footnote[5]{The self-energy is more accurately given by $\Delta E_{p} = \int_{0}^{\infty}\left[ \Delta E_{p}(\lambda) - \Delta E_{p}(0)\right]\, G(\lambda)  \, d\lambda.$ $^{\scriptsize{\cite{FeynmanQuantum}}}$})$ of a free electron (spin-$\frac{1}{2}$ case) with momentum $\bm{p}$ and bare mass $m$ that Feynman$^{\scriptsize{\cite{FeynmanQuantum}}}$ obtained was 
\begin{equation}\label{FeynmanSelfEnergy}
	\left(2\pi^2/e^{2} \right)E_{p} \Delta E_{p}= \big(m^{2}+ \frac{1}{2}\lambda^{2}\big) \int \frac{1}{(E_{f} + \omega)^{2} - E_{p}^{2}} \cdot \frac{E_{f}+ \omega}{E_{f}\ \omega}d\bm{k} + \frac{1}{2} \int \frac{d\bm{k}}{\omega}- \frac{1}{2} \int \frac{d\bm{k}}{E_{f}},
\end{equation}
where $E_{p}=\sqrt{\bm{p}^{2}+ m^{2}}$, $\omega=\sqrt{k^{2}+ \lambda^{2}}$, and $E_{f}=\sqrt{(\bm{p}+ \bm{k})^{2}+ m^{2}}$. \\
\\
The correction to mass that Feynman obtained as a result was
\begin{equation*}
	\Delta m = \frac{m\, e^{2}}{\pi}\left( \frac{3}{2}\ln \left(\frac{\lambda_{0}}{m} \right)+ \frac{3}{2}\right) \quad \text{where,} \quad \ln \lambda_{0} = \int_{0}^{\infty} G(\lambda)\ln \lambda\, d\lambda.  
\end{equation*}
In the limit as $\lambda_{0} \rightarrow \infty$, one recovers the result of usual electrodynamics where the mass correction diverges logarithmically. The mass correction term given above is also Lorentz invariant.\\
\\
However, the mass correction obtained from usual electrodynamics by employing the old-fashioned perturbation calculation will result in a term that is not Lorentz invariant.$^{\scriptsize{\cite{Schweber}}}$ Since formally the result diverges, the Lorentz non-invariant term wouldn't matter. However, if mass renormalization is to be performed unambiguously then the result, although formally divergent, should be Lorentz invariant. QED calculation techniques later developed by Feynman, Tomonaga, Schwinger, and Dyson do exactly this: the perturbation calculations are done in a manner that makes the Lorentz invariance apparent at every stage of the calculation.\\
\\
The corresponding expression for the self-energy of a spin-0 particle with momentum $\bm{p}$ and bare mass $m$ is 
\begin{equation}\label{Spin0SelfEnergy}
	\left(2\pi^2/e^{2} \right)E_{p} \Delta E_{p}= \big(m^{2}- \frac{1}{4}\lambda^{2}\big) \int \frac{1}{(E_{f} + \omega)^{2} - E_{p}^{2}} \cdot \frac{E_{f}+ \omega}{E_{f}\ \omega}d\bm{k} + \frac{1}{2} \int \frac{d\bm{k}}{\omega}+ \frac{1}{4} \int \frac{d\bm{k}}{E_{f}}
\end{equation}
There are two difficulties that must be overcome when carrying out self-energy calculation for the spin-0 case in the Hamiltonian method. The usual difficulty of conjugate momentum vanishing also persists in the modified electrodynamics.$^{\scriptsize{\cite{PodolskySchwed}}}$ Podolsky \& Schwed modified the Lagrangian so that conjugate momenta don't vanish but nonetheless lead to the same equations of motion for the fields. The other difficulty is using a gauge condition that is consistent with the commutation relations of the fields.\\
\\
The approach used in this paper follows the theory of constraints outlined by Dirac in his book ``Lectures on Quantum Mechanics".$^{\scriptsize{\cite{Dirac}}}$ Even with this approach, certain gauge choices, for example the Coulomb gauge $\nabla \cdot \bm{A} = 0$, cannot be used. Following Galvão \& Pimentel$^{\scriptsize{\cite{CarlosPimentel}}}$ a ``generalized Coulomb gauge", written as $(1-a^{2}\Box)\nabla \cdot \bm{A} = 0$, is instead used as the gauge condition on the fields. With this choice of gauge,
application of the theory of constraints leads to the following modified commutation relations for the fields $A_{\mu}^{(1)}$ and $A_{\mu}^{(2)}$ and their conjugate momenta:
\begin{align*}
	[A_{i}^{(1)}(\bm{x}), P_{j}^{(1)}(\bm{y})] &= i \hbar\, \delta_{ij} \delta^{3}(\bm{x}-\bm{y}) \\
	[A_{i}^{(2)}(\bm{x}), P_{j}^{(2)}(\bm{y})] &= -i \hbar\, \delta_{ij} \delta^{3}(\bm{x}-\bm{y}) - i \hbar\, \partial_{i}\partial_{j} \mathcal{G}
\end{align*}
where, $\mathcal{G}$ satisfies the partial differential equation $(1-a^{2}\nabla^{2})\nabla^{2} \mathcal{G} = \delta^{3}(\bm{x}-\bm{y})$ and $i,j=1,2,3$.\\
\\
The implication of these modified commutation relations is that, for the part $\bm{A}^{(1)}$ of the field everything is the same as it is for the usual (unmodified) electromagnetic field $\bm{A}$. However, for the part $\bm{A}^{(2)}$ things become different. First, the commutation relations of the creation and annihilation operators of this field become
\begin{equation*}
	[\hat{a}^{(2)}_{\lambda}(\bm{p}), \hat{a}^{\dagger (2)}_{\nu}(\bm{q})] = -\delta_{\lambda \nu} \delta^{3}(\bm{p}-\bm{q})
\end{equation*}
for the polarizations $\lambda, \nu = 1, 2$ perpendicular to $\bm{p}$, and 
\begin{equation*}
	[\hat{a}^{(2)}(\bm{p}), \hat{a}^{\dagger (2)}(\bm{q})] = -\frac{m^{2}}{E_{p}^{2}}\delta^{3}(\bm{p}-\bm{q})
\end{equation*}
for the polarization along the direction of $\bm{p}$. Second, the negative sign in the commutation relation means that the norm of some states will be negative. This is not desirable for a physical theory; it results in negative probabilities, which is nonsensical. In order to avoid this difficulty, one must define the action of the operators $\hat{a}^{(2)}_{\lambda}(\bm{p})$ and $\hat{a}^{\dagger (2)}_{\lambda}(\bm{p})$ on the state vectors $|n, \lambda \rangle$ ($ |n, \lambda \rangle$ are eigenstates of the number operator $N_{\lambda}=-\hat{a}^{\dagger (2)}_{\lambda}\hat{a}^{(2)}_{\lambda}$) such that:
\begin{align*}
	\hat{a}^{(2)}_{\lambda}| n,\lambda \rangle &= -\sqrt{n}\, |n-1\rangle \quad \text{and}\\
	\hat{a}^{\dagger(2)}_{\lambda}| n,\lambda \rangle &= \sqrt{n+1}\, |n+1\rangle,
\end{align*}
and also redefine the inner product as $\langle n |\mathcal{O} | n \rangle$ so that the inner product comes out positive for all states $| n \rangle$. Here, $\mathcal{O}$ is an operator that has the following properties:
\begin{enumerate}
	\item $\mathcal{O}| n \rangle = (-1)^{n}| n \rangle$,
	\item $\mathcal{O}^{\dagger}= \mathcal{O}$,
	\item $\mathcal{O}\hat{a}^{(2)}= - \hat{a}^{(2)}\mathcal{O}$, \quad \text{and}
	\item $\mathcal{O}\hat{a}^{\dagger (2)}= - \hat{a}^{\dagger (2)}\mathcal{O}$.
\end{enumerate}
This way of avoiding the difficulty of negative norm was first carried out by Gupta$^{\scriptsize{\cite{Gupta}}}$ and independently by Bleuler in 1950.\\
\\
After sorting through these difficulties and applying of the ``old fashioned perturbation" method one obtains the self-energy expression \eqref{Spin0SelfEnergy} for a spin-0 particle interacting with the electromagnetic field.
Thus, Feynman's modification \eqref{FeynmanMod} (or equivalently the modification suggested by Podolsky \& Schwed in the Hamiltonian scheme) for the spin-$\frac{1}{2}$ case applies almost without any change to the spin-0 case as well.
\bibliography{references_paper1}

\end{document}